# A Workload and Programming Ease Driven Perspective of Processing-in-Memory

Saugata Ghose[†]    Amirali Boroumand[†]    Jeremie S. Kim[†§]    Juan Gómez-Luna[§]    Onur Mutlu[§†]

[†]*Carnegie Mellon University*    [§]*ETH Zürich*

*Many modern and emerging applications must process increasingly large volumes of data. Unfortunately, prevalent computing paradigms are not designed to efficiently handle such large-scale data: the energy and performance costs to move this data between the memory subsystem and the CPU now dominate the total costs of computation. This forces system architects and designers to fundamentally rethink how to design computers.* Processing-in-memory *(PIM) is a computing paradigm that avoids most data movement costs by* bringing computation to the data. *New opportunities in modern memory systems are enabling architectures that can perform varying degrees of processing inside the memory subsystem. However, there are many practical system-level issues that must be tackled to construct PIM architectures, including enabling workloads and programmers to easily take advantage of PIM. This article examines three key domains of work towards the practical construction and widespread adoption of PIM architectures. First, we describe our work on systematically identifying opportunities for PIM in real applications, and quantify potential gains for popular emerging applications (e.g., machine learning, data analytics, genome analysis). Second, we aim to solve several key issues on programming these applications for PIM architectures. Third, we describe challenges that remain for the widespread adoption of PIM.*

## 1. Introduction

A wide range of application domains have emerged as computing platforms of all types have become more ubiquitous in society. Many of these modern and emerging applications must now process very large data sets [1-8]. As an example, an object classification algorithm in an augmented reality application typically trains on millions of example images and video clips, and performs classification on real-time high-definition video streams [7, 9]. In order to process meaningful information from the large amounts of data, applications turn to artificial intelligence (AI), or machine learning (ML), and data analytics to methodically mine through the data and extract key properties about the data set.

Due to the increasing reliance on manipulating and mining through large sets of data, these modern applications greatly overwhelm the data storage and movement resources of a modern computer. In a contemporary computer, the main memory (consisting of DRAM) is not capable of performing any operations on data. As a result, to perform *any* operation on data that is stored in memory, the data needs to be *moved* from the memory to the CPU via the *memory channel*, a pin-limited off-chip bus (e.g., conventional double data rate, or DDR, memories make use of a 64-bit memory channel [10-12]). To move the data, the CPU must issue a request to the memory controller, which then issues commands across the *memory channel* to the DRAM module containing the data. The DRAM module then reads and returns the data across the memory channel, and the data moves through the cache hierarchy before being stored in a CPU cache. The CPU can operate on the data only once the data is loaded from the cache into a CPU register.

Unfortunately, for modern and emerging applications, the large amounts of data that need to move across the memory channel create a large *data movement bottleneck* in the computing system [13-14]. The data movement bottleneck incurs a heavy penalty in terms of both performance and energy consumption [13-20]. First, there is a long latency and significant energy involved in bringing data from DRAM. Second, it is difficult to send a large number of requests to memory in parallel, in part because of the narrow width of the memory channel. Third, despite the costs of bringing data into memory, much of this data is not reused by the CPU, rendering the caching either highly inefficient or completely unnecessary [5, 21], especially for modern workloads with very large datasets and random access patterns. Today, the total cost of computation, in terms of performance and in terms of energy, is dominated by the cost of data movement for modern data-intensive workloads such as machine learning and data analytics [5, 15, 16, 21-25].

The high cost of data movement is forcing architects to rethink the fundamental design of computer systems. As data-intensive applications become more prevalent, there is a need to *bring computation closer to the data*, instead of moving data across the system to distant compute units. Recent advances in memory design enable the opportunity for architects to avoid unnecessary data movement by performing *processing-in-memory* (PIM), also known as *near-data processing* (NDP). The idea of performing PIM has been proposed for at least four decades [26-36], but earlier efforts were *not* widely adopted due to the difficulty of integrating processing elements for computation with DRAM. Innovations such as (1) 3D-stacked memory dies that combine a logic layer with DRAM layers [5, 37-40], (2) the ability to perform logic operations using memory cells themselves inside a memory chip [18, 20, 41-49], and (3) the emergence of potentially more computation-friendly resistive memory technologies [50-61] provide new opportunities to embed general-purpose computation *directly within the memory* [5, 16-19, 21, 22, 24, 25, 41-43, 47-49, 62-100].

While PIM can allow many data-intensive applications to avoid moving data from memory to the CPU, it introduces new challenges for system architects and programmers. In this work, we examine two major areas of challenges, and discuss solutions that we have developed for each challenge. First, programmers need to be able to *identify opportunities in their applications where PIM can improve their target objectives* (e.g., application performance, energy consumption). As we discuss in Section 3, the decision on whether to execute part or all of an application in memory depends on (1) architectural constraints, such as area and energy limitations, and the type of logic implementable within memory; and (2) application properties, such as the intensities of computation and memory accesses, and the amount of data shared across different functions. To solve this first challenge, we have developed toolflows that help the programmer to systematically determine how to partition work between *PIM logic* (i.e., processing elements on the memory side) and the CPU, in order to meet all architectural design constraints and maximize targeted benefits [16, 22-24, 75]. Second, system architects and programmers must establish *efficient interfaces and mechanisms that allow programs to easily take advantage of the benefits of PIM*. In particular, the processing logic inside memory does *not* have quick access to important mechanisms required by modern programs and systems, such as cache coherence and address translation, which programmers rely on for software development productivity. To solve this second challenge, we develop a series of interfaces and mechanisms that are designed specifically to allow programmers to use PIM in a way that preserves conventional programming models [5, 16-24, 62, 75].

In providing a series of solutions to these two major challenges, we tackle many of the fundamental barriers that have prevented PIM from being adopted widely, in a programmer-friendly way. We find that a number of future challenges remain against the adoption of PIM, and we discuss them briefly in Section 6. We hope that our work inspires researchers to address these and other future challenges, and that both our work and future works help to enable the widespread commercialization and usage of PIM-based computing systems.

## 2. Overview of Processing-in-Memory (PIM)

The costs of data movement in an application continue to increase significantly as applications process larger data sets. Processing-in-memory provides a viable path to eliminate unnecessary data movement, by bringing part or all of the computation into the memory. In this section, we briefly examine key enabling technologies behind PIM, and how new advances and opportunities in memory design have brought PIM significantly closer to realization.

### 2.1. The Initial Push for PIM

Proposals for PIM architectures extend back as far as the 1960s. Stone's Logic-in-Memory computer is one of the earliest PIM architectures, in which a distributed array of memories combines small processing elements with small amounts of RAM to perform computation within the memory array [36]. Between the 1970s and the early 2000s, a number of subsequent works propose different ways to integrate computation and memory, which we broadly categorize into two families of work. In the first family, which includes NON-VON [35], Computational RAM [27, 28], EXECUBE [31], Terasys [29], and IRAM [34], architects add logic within DRAM to perform data-parallel operations. In the second family of works, such as Active Pages [33], FlexRAM [30], Smart Memories [32], and DIVA [26], architects propose more versatile substrates that tightly integrate logic and reconfigurability within DRAM itself to increase flexibility and the available compute power. Unfortunately, many of these works were hindered by the limitations of existing memory technologies, which prevented the practical integration of logic in or near the memory.

### 2.2. New Opportunities in Modern Memory Systems

Due to the increasing need for large memory systems by modern applications, DRAM scaling is being pushed to its practical limits [101-104]. It is becoming more difficult to increase the density [101, 105-107], reduce the latency [107-112], and decrease the energy consumption [101, 113, 114] of conventional DRAM architectures. In response, memory manufacturers are actively developing two new approaches for main memory system design, both of which can be exploited to overcome prior barriers to implementing PIM architectures.

The first major innovation is *3D-stacked memory* [5, 37-40]. In a 3D-stacked memory, multiple layers of memory (typically DRAM) are stacked on top of each other, as shown in **Figure 1**. These layers are connected together using vertical *through-silicon vias* (TSVs) [38, 39]. With current manufacturing process technologies, thousands of TSVs can be placed within a single 3D-stacked memory chip. The TSVs provide much greater internal memory bandwidth than the narrow memory channel. Examples of 3D-stacked DRAM available commercially include High-Bandwidth Memory (HBM) [37, 38], Wide I/O [115], Wide I/O 2 [116], and the Hybrid Memory Cube (HMC) [40].

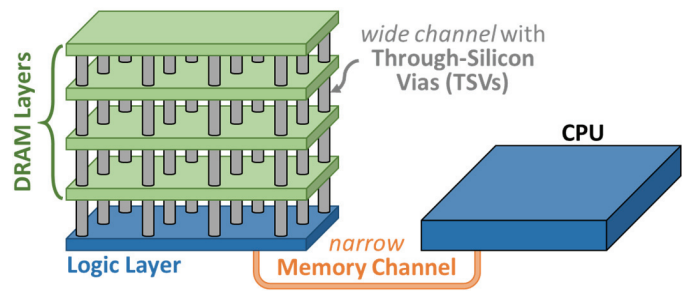

**Figure 1** High-level overview of a 3D-stacked DRAM architecture. Reproduced from [14].

In addition to the multiple layers of DRAM, a number of prominent 3D-stacked DRAM architectures, including HBM and HMC, incorporate a *logic layer* inside the chip [37, 38, 40]. The logic layer is typically the bottommost layer of the chip, and is connected to the same TSVs as the memory layers. The logic layer provides a space inside the DRAM chip where architects can implement functionality that interacts with both the processor and the DRAM cells. Currently, manufacturers make limited use of the logic layer, presenting an opportunity for architects to implement new PIM logic in the available area of the logic layer. We can potentially add a wide range of computational logic (e.g., general-purpose cores, accelerators, reconfigurable architectures) in the logic layer, as long as the added logic meets area, energy, and thermal dissipation constraints.

The second major innovation is the use of byte-addressable resistive *nonvolatile memory* (NVM) for the main memory subsystem. In order to avoid DRAM scaling limitations entirely, researchers and manufacturers are developing new memory devices that can store data at much higher densities than the typical density available in existing DRAM manufacturing process technologies. Manufacturers are exploring at least three types of emerging NVMs to augment or replace DRAM at the main memory layer: (1) *phase-change memory* (PCM) [50-56], (2) *magnetic RAM* (MRAM) [57, 58], and (3) *metal-oxide resistive RAM* (RRAM) or *memristors* [59-61]. All three of these NVM types are expected to provide memory access latencies and energy usage that are competitive with or close enough to DRAM, while enabling much larger capacities per chip and nonvolatility in main memory.

NVMs present architects with an opportunity to redesign how the memory subsystem operates. While it can be difficult to modify the design of DRAM arrays due to the delicacy of DRAM manufacturing process technologies as we approach scaling limitations, NVMs have yet to approach such scaling limitations. As a result, architects can potentially design NVM memory arrays that integrate PIM functionality. A promising direction for this functionality is the ability to manipulate NVM cells at the circuit level in order to perform logic operations using the memory cells themselves. A number of recent works have demonstrated that NVM cells can be used to perform a complete family of Boolean logic operations [41-46], similar to such operations that can be performed in DRAM cells [18, 20, 47-49].

### 2.3. Two Approaches: Processing-Near-Memory vs. Processing-Using-Memory

Many recent works take advantage of the memory technology innovations that we discuss in Section 2.2 to enable PIM. We find that these works generally take one of two approaches, which are summarized in **Table 1**: (1) *processing-near-memory* or (2) *processing-using-memory*. Processing-*near*-memory involves adding or integrating PIM logic (e.g., accelerators, very small in-order cores, reconfigurable logic) close to or inside the memory (e.g., [5, 6, 16, 21-25, 62, 64-73, 75, 77, 79, 81, 83-87, 90, 91, 117]). Many of these works place PIM logic inside the logic layer of 3D-stacked memories or at the memory controller, but recent advances in silicon interposers (in-package wires that connect directly to the through-silicon vias in a 3D-stacked chip) also allow for separate logic chips to be placed in the same die package as a 3D-stacked memory while still taking advantage of the TSV bandwidth. In contrast, processing-*using*-memory makes use of intrinsic properties and operational principles of the memory cells and cell arrays themselves, by inducing interactions between cells such that the cells and/or cell arrays can perform computation. Prior works show that processing-using-memory is possible using static RAM (SRAM) [63, 76, 100], DRAM [17-20, 47-49, 80, 87, 118], PCM [41], MRAM [44-46], or RRAM/memristive [42, 43, 88, 92-99] devices. Processing-using-memory architectures enable a range of different functions, such as bulk copy and data initialization [17, 19, 63], bulk bitwise operations (e.g., a complete set of Boolean logic operations) [18, 41, 44-49, 63, 80, 106-108, 118], and simple arithmetic operations (e.g., addition, multiplication, implication) [42, 43, 63, 76, 80, 88, 92-100].

**Table 1** Summary of enabling technologies for the two approaches to PIM used by recent works.

| Approach | Enabling Technologies |
|---|---|
| Processing-Near-Memory | Logic layers in 3D-stacked memory<br>Silicon interposers<br>Logic in memory controllers |
| Processing-Using-Memory | SRAM<br>DRAM<br>Phase-change memory (PCM)<br>Magnetic RAM (MRAM)<br>Resistive RAM (RRAM)/memristors |

### 2.4. Challenges to the Adoption of PIM

In order to build PIM architectures that are adopted and readily usable by most programmers, there are a number of challenges that need to be addressed. In this work, we discuss two of the most significant challenges facing PIM. First, programmers need to be able to identify what portions of an application are suitable for PIM, and architects need to understand the constraints imposed by different substrates when designing PIM logic. We address this challenge in Section 3. Second, once opportunities for PIM have been identified and PIM architectures have been designed, programmers need a way to extract the benefits of PIM without having to resort to complex programming models. We address this challenge in Section 4. While these two challenges represent some of the largest obstacles to widespread adoption for PIM, a number of other important challenges remain, which we discuss briefly in Section 6.

### 3. Identifying Opportunities for PIM in Applications

In order to decide when to use PIM, we must first understand which types of computation can benefit from being moved to memory. The opportunities for an application to benefit from PIM depend on (1) the constraints of the target architecture, and (2) the properties of the application.

### 3.1. Design Constraints for PIM

The target architecture places a number of fundamental constraints on the types of computation that can benefit from PIM. As we discuss in Section 2.3, there are two approaches to implementing PIM (processing-near-memory and processing-using-memory). Each approach has its own constraints on what type of logic can be efficiently and effectively implemented in memory.

In the case of *processing-near-memory*, PIM logic must be added close to the memory, either in the logic layer of a 3D-stacked memory chip or in the same package. This places a limit on how much PIM logic can be added. For example, in an HMC-like 3D-stacked memory architecture implemented using a 22 nm manufacturing processing technology, we estimate that there is around 50–60 mm$^2$ of area available for architects to add new logic into the DRAM logic layer [40]. The available area can be further limited by the architecture of the memory. For example, in HMC, the 3D-stacked memory is partitioned into multiple *vaults* [40], which are vertical slices of 3D-stacked DRAM. Logic placed in a vault has fast access to data stored in the memory layers of the same vault, as the logic is directly connected to the memory in the vault by the TSVs (see Section 2.2), but accessing data stored in a different vault takes significantly longer latency. As a result, architects often replicate PIM logic in each vault, to minimize the latency of PIM operations. The trade-off of this is that the amount of area available *per vault* is significantly lower: for a 32-vault 3D-stacked memory chip, there is approximately 3.5–4.4 mm$^2$ of area available for PIM logic [119-121].

A number of target computing platforms have additional constraints beyond area. For example, consumer devices such as smartphones, tablets, and netbooks are extremely stringent in terms of *both* the area and energy budget they can accommodate for any new hardware enhancement. Any additional logic added to memory can potentially translate into a significant cost in consumer devices. In fact, unlike PIM logic that is added to server or desktop environments, consumer devices may not be able to afford the addition of full-blown general-purpose PIM cores [22-24, 68, 120], GPU PIM cores [75, 85, 90], or complex PIM accelerators [5, 62, 119] to 3D-stacked memory. As a result, a major challenge for enabling PIM in consumer devices is to identify what kind of in-memory logic can both (1) maximize energy efficiency and (2) be implemented at minimum possible cost. Another constraint is thermal dissipation in 3D-stacked memory, as adding PIM logic in the logic layer can potentially raise the DRAM temperature beyond acceptable levels [85, 90].

In the case of *processing-using-memory*, the cells and memory array themselves are used to implement PIM logic. Additional logic in the controller and/or in the array itself may be required to enable logic operations on the cells or in the memory array, or to provide more specialized functionality beyond what the cells and memory array themselves can perform easily (e.g., dedicated adders or shifters).

### 3.2. Choosing What to Execute in Memory

After the constraints on what type of hardware can potentially be implemented in memory are determined, the properties of the application itself are a key indicator of whether portions of an application benefit from PIM. A naïve assumption may be to move highly-memory-intensive applications completely to PIM logic. However, we find that there are cases where portions of these applications still benefit from remaining on the CPU. For example, many proposals for PIM architectures add small general-purpose cores near memory (which we call *PIM cores*). While PIM cores tend to be ISA-compatible with the CPU, and can execute any part of the application, they cannot afford to have large, multi-level cache hierarchies or execution logic that is as complex as the CPU, due to area, energy, and thermal constraints. PIM cores often have no or small caches, restricting the amount of temporal locality they can exploit, and no sophisticated aggressive out-of-order or superscalar execution logic, limiting the PIM cores' abilities to extract instruction-level parallelism (ILP). As a result, portions of an application that are either (1) compute-intensive or (2) cache-friendly should remain on the larger, more sophisticated CPU cores [16, 21-24, 75].

We find that in light of these constraints, it is important to identify which *portions of an application are suitable for PIM*. We call such portions *PIM targets*. While PIM targets can be identified manually by a programmer, the identification would require significant programmer effort along with a detailed understanding of the hardware trade-offs between CPU cores and PIM cores. For architects who are adding custom PIM logic (e.g., fixed-function accelerators, which we call *PIM accelerators*) to memory, the trade-offs between CPU cores and PIM accelerators may not be known before determining which portions of the application are PIM targets, since the PIM accelerators are tailored for the PIM targets.

To alleviate the burden of manually identifying PIM targets, we develop a systematic toolflow for identifying PIM targets in an application [16, 22-24]. This toolflow uses a system that executes the entire application on the CPU to evaluate whether each PIM target meets the constraints of the system under consideration. For example, when we evaluate workloads for consumer devices, we use hardware performance counters and our energy model to identify candidate functions that could be PIM targets. A function is a PIM target candidate in a consumer device if (1) it consumes the most energy out of all functions in the workload, since energy reduction is a primary objective in consumer workloads; (2) its data movement consumes a significant fraction (e.g., more than 20%) of the total workload energy, to maximize the potential energy benefits of offloading to PIM; (3) it is memory-intensive (i.e., its last-level cache *misses per kilo instruction*, or MPKI, is greater than 10 [122-125]), as the energy savings of PIM is higher when more data movement is eliminated; and (4) data movement is the single largest component of the function's energy consumption. We then check if each candidate function is amenable to PIM logic implementation using two criteria. First, we discard any PIM targets that incur any performance loss when run on simple PIM logic (i.e., PIM core, PIM accelerator). Second, we discard any PIM targets that require more area than is available in the logic layer of 3D-stacked memory. Note that for pre-built PIM architectures with fixed PIM logic, we instead discard any PIM targets that cannot be executed on the existing PIM logic.

While our toolflow was initially designed to identify PIM targets for consumer devices [16], the toolflow can be modified to accommodate any other hardware constraints. For example, in our work on reducing the cost of cache coherence in PIM architectures [22-24], we consider the amount of *data sharing* (i.e., the total number of cache lines that are read concurrently by the CPU and by PIM logic). In that work, we eliminate any potential PIM target that would result in a high amount of data sharing if the target were offloaded to a PIM core, as this would induce a large amount of cache coherence traffic between the CPU and PIM logic that would counteract the data movement savings (see Section 4.2).

### 3.3. Case Study: PIM Opportunities in TensorFlow

By performing our constraint analysis (Section 3.1) and using our systematic PIM target toolflow (Section 3.2), we find that a number of key modern workloads are well-suited for PIM. In particular, we find that machine learning and data analytics workloads are particularly amenable for PIM, as they are often partitioned into compute-intensive and memory-intensive application phases. These workloads benefit highly from PIM when only the memory-intensive PIM targets (that fit our system constraints) are offloaded to PIM logic. Such workloads include neural network inference [126], graph analytics [127-132], and hybrid transactional/analytical processing databases [133-135].

As a case study, we present a detailed analysis using our PIM target identification approach for TensorFlow Lite [126], a version of Google's TensorFlow machine learning library that is specifically tailored for mobile and embedded platforms. TensorFlow Lite enables a variety of tasks, such as image classification, face recognition, and Google Translate's instant visual translation [136], all of which perform inference on consumer devices using a convolutional neural network that was pre-trained on cloud servers. We target a processing-near-memory platform in this case study, where we add small in-order PIM cores or fixed-function PIM accelerators into the logic layer of a 3D-stacked DRAM. We model a 3D-stacked DRAM similar to the Hybrid Memory Cube [40], where the memory contains sixteen *vaults* (i.e., vertical slices of DRAM). We add one PIM core or PIM accelerator per vault, ensuring that the area of the PIM core or the PIM accelerator does not exceed the total available area for logic inside each vault (3.5–4.4 mm$^2$ [119-121]). Each PIM core or PIM accelerator can execute one PIM target at a time. Details about our methodology, along with the specific parameters of the target platform, can be found in our prior work [16].

Inference begins by feeding input data (e.g., an image) to a neural network. A neural network is a directed acyclic graph consisting of multiple layers. Each layer performs a number of calculations and forwards the results to the next layer. The calculation can differ for each layer, depending on the type of the layer. A fully-connected layer performs matrix multiplication (MatMul) on the input data, to extract high-level features. A 2-D convolution layer applies a convolution filter (Conv2D) across the input data, to extract low-level features. The last layer of a neural network is the output layer, which performs classification to generate a prediction based on the input data.

**Energy Analysis: Figure 2** shows the breakdown of the energy consumed by each function in TensorFlow Lite, for four different input networks: ResNet-v2-152 [137], VGG-19 [138], Residual-GRU [139], and Inception-ResNet-v2 [140]. As convolutional neural networks (CNNs) consist mainly of 2-D convolution layers and fully-connected layers [141], the majority of energy is spent on these two types of layers. However, we find that there are two other functions that consume a significant fraction of the system energy: packing/unpacking and quantization. Packing and unpacking reorder the elements of matrices to minimize cache misses during matrix multiplication. Quantization converts 32-bit floating point and integer values (used to represent both the weights and activations of a neural network) into 8-bit integers, which improves the execution time and energy consumption of inference by reducing the complexity of operations that the CPU needs to perform. These two functions together account for 39.3% of total system energy on average. The rest of the energy is spent on a variety of other functions such as random sampling, reductions, and simple arithmetic, each of which contributes to less than 1% of total energy consumption (labeled *Other* in Figure 2).

Even though the main goal of packing and quantization is to reduce energy consumption and inference latency, our analysis shows that they generate a large amount of data movement, and thus, lose part of the energy savings they aim to achieve. **Figure 3** shows that a significant portion (27.4% on average) of the execution time is spent on the packing and quantization process. We do not consider Conv2D and MatMul as being candidates for offloading to PIM logic because (1) a majority (67.5%) of their energy is spent on computation; and (2) Conv2D and MatMul require a relatively large and sophisticated amount of PIM logic [77, 119], which may not be cost-effective for consumer devices.

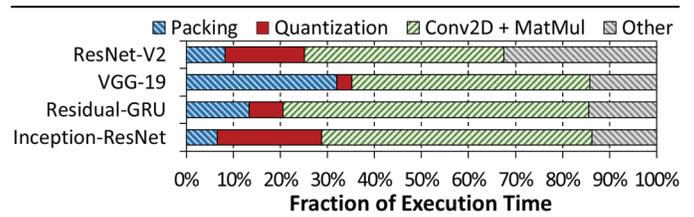

**Figure 3** Execution time breakdown of inference. Reproduced from [16].

**PIM Effectiveness for Packing:** We highlight how PIM can be used to effectively improve the performance and energy consumption of packing. GEneralized Matrix Multiplication (GEMM) is the core building block of neural networks, and is used by both 2-D convolution and fully-connected layers. These two layers account for the majority of TensorFlow Lite execution time. To implement fast and energy-efficient GEMM, TensorFlow Lite employs a low-precision, quantized GEMM library called *gemmlowp* [142]. The gemmlowp library performs GEMM by executing its innermost kernel, an architecture-specific GEMM code portion for small fixed-size matrix chunks, multiple times. First, gemmlowp fetches matrix chunks which fit into the LLC from DRAM. Then, it executes the GEMM kernel on the fetched matrix chunks in a block-wise manner.

Each *GEMM operation* (i.e., a single matrix multiply calculation using the gemmlowp library) involves three steps. First, to minimize cache misses, gemmlowp employs a process called *packing*, which reorders the matrix chunks based on the memory access pattern of the kernel to make the chunks cache-friendly. Second, the actual GEMM computation (i.e., the innermost GEMM kernel) is performed. Third, after performing the computation, gemmlowp performs *unpacking*, which converts the result matrix chunk back to its original order.

Packing and unpacking account for up to 40% of the total system energy and 31% of the inference execution time, as shown in Figures 2 and 3, respectively. Due to their unfriendly cache access pattern and the large matrix sizes, packing and unpacking generate a significant amount of data movement. For instance, for VGG-19, 35.3% of the total energy goes to data movement incurred by packing-related functions. On average, we find that data movement is responsible for 82.1% of the total energy consumed during the packing/unpacking process, indicating that packing and unpacking are bottlenecked by data movement.

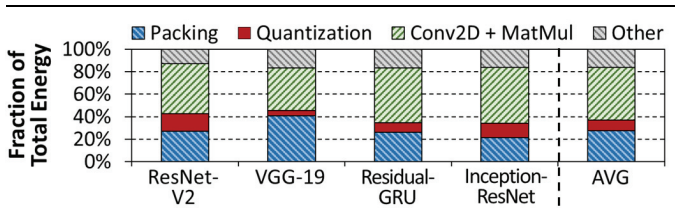

**Figure 2** Energy breakdown during TensorFlow Lite inference execution on four input networks. Reproduced from [16].

Packing and unpacking are simply pre-processing steps, to prepare data in the right format for the innermost GEMM kernel. Ideally, the CPU should execute only the innermost GEMM kernel, and assume that packing and unpacking are already taken care of. PIM can enable such a scenario by performing packing and unpacking without *any* CPU involvement. Our PIM logic packs matrix chunks, and sends the packed chunks to the CPU, which executes the innermost GEMM kernel. Once the innermost GEMM kernel completes, the PIM logic receives the result matrix chunk from the CPU, and unpacks the chunk while the CPU executes the innermost GEMM kernel on a different matrix chunk.

**PIM Effectiveness for Quantization:** TensorFlow Lite performs quantization twice for each Conv2D operation. First, quantization is performed on the 32-bit input matrix before Conv2D starts, which reduces the complexity of operations required to perform Conv2D on the CPU by reducing the width of each matrix element to 8 bits. Then, Conv2D runs, during which gemmlowp generates a 32-bit result matrix. Quantization is performed for the second time on this result matrix (this step is referred to as *re-quantization*). Accordingly, invoking Conv2D more frequently (which occurs when there are more 2-D convolution layers in a network) leads to higher quantization overheads. For example, VGG-19 requires only 19 Conv2D operations, incurring small quantization overheads. On the other hand, ResNet-v2 requires 156 Conv2D operations, causing quantization to consume 16.1% of the total system energy and 16.8% of the execution time. The quantization overheads are expected to increase as neural networks get deeper, as a deeper network requires a larger matrix.

**Figure 4a** shows how TensorFlow quantizes the result matrix using the CPU. First, the entire matrix needs to be scanned to identify the minimum and maximum values of the matrix (❶ in the figure). Then, using the minimum and maximum values, the matrix is scanned a second time to convert each 32-bit element of the matrix into an 8-bit integer (❷). These steps are repeated for re-quantization of the result matrix (❸ and ❹). The majority of the quantization overhead comes from data movement. Because both the input matrix quantization and the result matrix re-quantization need to scan a large matrix twice, they exhibit poor cache locality and incur a large amount of data movement. For example, for the ResNet-v2 network, 73.5% of the energy consumed during quantization is spent on data movement, indicating that the computation is relatively cheap (in comparison, only 32.5% of Conv2D/MatMul energy goes to data movement, while the majority goes to multiply–accumulate computation). 19.8% of the total data movement energy of inference execution comes from quantization and re-quantization. As **Figure 4b** shows, we can offload both quantization (❺ in the figure) and re-quantization (❻) to PIM to eliminate data movement. This frees up the CPU to focus on GEMM calculation, and allows the next Conv2D operation to be performed in parallel with re-quantization (❼).

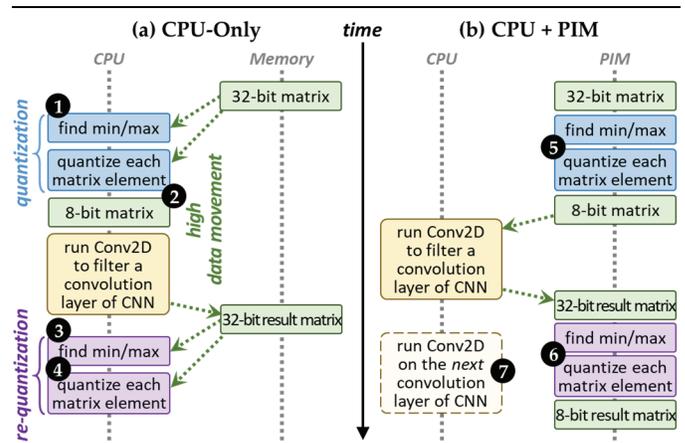

**Figure 4** Quantization on (a) CPU vs. (b) PIM. Reproduced from [16].

**Evaluation:** We evaluate how TensorFlow Lite benefits from PIM execution using (1) custom 64-bit low power single-issue cores similar in design to the ARM Cortex-R8 [143]; and (2) fixed-function PIM accelerators designed for packing and quantization operations, with each accelerator consisting of four simple ALUs and consuming less than 0.25 mm$^2$ of area [16]. **Figure 5** (left) shows the energy consumption of PIM execution using PIM cores (*PIM-Core*) or fixed-function PIM accelerators (*PIM-Acc*) for the four most time- and energy-consuming GEMM operations for each input neural network in packing and quantization, normalized to a processor-only baseline (*CPU-Only*). We make three key observations. First, PIM-Core and PIM-Acc decrease the total energy consumption of a consumer device system by 50.9% and 54.9%, on average across all four input networks, compared to CPU-Only. Second, the majority of the energy savings comes from the large reduction in data movement, as the computation energy accounts for a negligible portion of the total energy consumption. For instance, 82.6% of the energy reduction for packing is due to the reduced data movement. Third, we find that the data-intensive nature of these kernels and their low computational complexity limit the energy benefits PIM-Acc provides over PIM-Core.

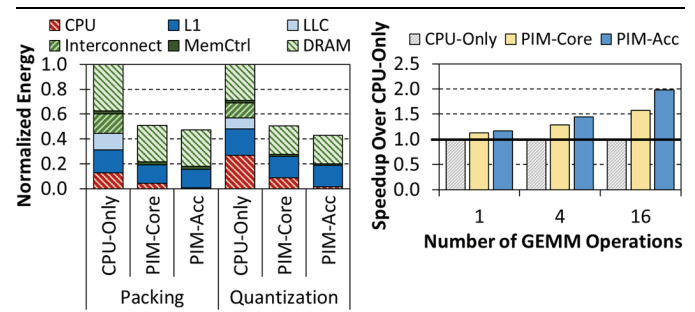

**Figure 5** Energy (left) and performance (right) for TensorFlow Lite kernels, averaged across four neural network inputs: ResNet-v2 [137], VGG-19 [138], Residual-GRU [139], Inception-ResNet [140]. Adapted from [16].

Figure 5 (right) shows the speedup of PIM-Core and PIM-Acc over CPU-Only as we vary the number of GEMM operations performed. For CPU-Only, we evaluate a scenario where the CPU performs packing, GEMM calculation, quantization, and unpacking. To evaluate PIM-Core and PIM-Acc, we assume that packing and quantization are handled by the PIM logic, and the CPU performs GEMM calculation. We find that, as the number of GEMM operations increases, PIM-Core and PIM-Acc provide greater performance improvements over CPU-Only. For example, for one GEMM operation, PIM-Core and PIM-Acc achieve speedups of 13.1% and 17.2%, respectively. For 16 GEMM operations, the speedups of PIM-Core and PIM-Acc increase to 57.2% and 98.1%, respectively, over CPU-Only. These improvements are the result of PIM logic (1) exploiting the higher bandwidth and lower latency of 3D-stacked memory, and (2) enabling the CPU to perform GEMM in parallel while the PIM logic handles packing and quantization.

We conclude that our approach to identifying PIM targets can be used to significantly improve performance and reduce energy consumption for the TensorFlow Lite mobile machine learning framework.

## 4. Programming PIM Architectures: Key Issues

While many applications have significant potential to benefit from PIM, a number of practical considerations need to be made with regards to *how* portions of an application are offloaded, and how this offloading can be accomplished without placing an undue burden on the programmer. When a portion of an application is offloaded to PIM logic, the PIM logic executes the offloaded piece of code, which we refer to as a *PIM kernel*. In this section, we study four key issues that affect the programmability of PIM architectures: (1) the different granularities of an offloaded PIM kernel, (2) how to handle data sharing between PIM kernels and CPU threads, (3) how to efficiently provide PIM kernels with access to essential virtual memory address translation mechanisms, and (4) how to automate the identification and offloading of PIM targets (i.e., portions of an application that are suitable for PIM; see Section 3.2).

### 4.1. Offloading Granularity

In Section 3.3, our case study on identifying opportunities for PIM in TensorFlow Lite makes an important assumption: PIM kernels are offloaded at the granularity of an *entire function*. However, there are a number of different granularities at which PIM kernels can be offloaded. Each granularity requires a different interface and different design decisions. We evaluate four offloading granularities in this section: (1) a single instruction, (2) a bulk operation, (3) an entire function, and (4) an entire application.

At one extreme, a PIM kernel can consist of a *single instruction* from the view of the CPU. For example, a PIM-enabled instruction (PEI) [21] can be added to an existing ISA, where each PIM operation is expressed and semantically operates as a single instruction. **Figure 6** shows an example architecture that can be used to enable PEIs [21]. In this architecture, a PEI is executed on a PEI Computation Unit (PCU). To enable PEI execution in either the host CPU or in memory, a PCU is added to each host CPU and to each vault in an HMC-like 3D-stacked memory. While the work done in a PCU for a PEI might have required multiple CPU instructions in the baseline CPU-only architecture, the CPU only needs to execute a single PEI instruction, which is sent to a central PEI Management Unit (PMU in Figure 6). The PMU launches the appropriate PIM operation on one of the PCUs. Implementing PEIs with low complexity and minimal changes to the system requires three key rules. First, for every PIM operation, there is a single PEI in the ISA that is used by the CPU to trigger the operation. This keeps the mapping between PEIs and PIM operations simple, and allows for the gradual introduction of new instructions. This also avoids the need for virtual memory address translation in memory, as the translation is done in the CPU before sending the PEI to memory. Second, a PIM operation is limited to operating on a single cache line. This (1) eliminates the need for careful data mapping, by ensuring that a single PEI operates on data that is mapped to a single memory controller; and (2) eases cache coherence, by needing no more than a single cache line to be kept coherent between a PEI performed in memory and the CPU cache. Third, a PEI is treated as atomic with respect to other PEIs, and uses memory fences to enforce atomicity between a PEI and a normal CPU instruction. An architecture with support for PEIs increases the average performance across ten graph processing, machine learning, and data mining applications by 32% over a CPU-only baseline for small input sets, and by 47% for large input sets [21]. While PEIs allow for simple coordination between CPU threads and PIM kernels, they limit both the complexity of computation performed and the amount of data processed by any one PIM kernel, which can incur high overheads when a large number of PIM operations need to be performed.

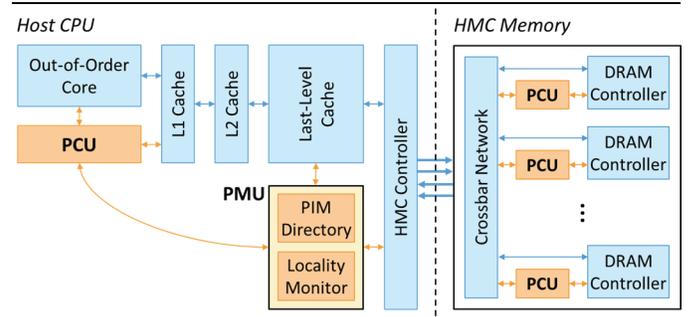

**Figure 6** Example architecture for PIM-enabled instructions. Adapted from [21].

One approach to perform more work than a single PEI is to offload *bulk operations* to memory, as is done by a number of processing-using-memory architectures. For a bulk operation, the same operation is performed across a large, often contiguous, region of memory (e.g., an 8KB row of DRAM). Mechanisms for processing-using-memory can perform a variety of bulk functions, such as bulk copy and data initialization [17, 19], bulk bitwise operations [18, 20, 47-49, 80], and simple arithmetic operations [42, 43, 63, 76, 88, 92-100]. As a representative example, the Ambit processing-using-memory architecture, which enables bulk bitwise operations using DRAM cells, accelerates the performance of database queries and operations on the *set* data structure by 3x–12x over a CPU-only baseline [18]. There are two trade-offs to performing bulk operations in memory. First, there are limitations in the amount of data that a single bulk operation processes: for example, a bulk bitwise operation in Ambit *cannot* be performed on less than one row at a time. Second, the operations that a processing-using-memory architecture can perform are much simpler than those that a general-purpose core can perform, due to limits in the amount of logical functionality that can be implemented inside the memory array.

A second approach to perform more work than a single PEI is to offload at the granularity of an application *function* or a block of instructions in the application [16, 22-24, 62, 75]. There are several ways to demarcate which portions of an application should be offloaded to PIM. One approach is to surround the portion with compiler directives. For example, if we want to offload a function to PIM, we can surround it with `#PIM_begin` and `#PIM_end` directives, which a compiler can use to generate a thread for execution on PIM. This approach requires compiler and/or library support to dispatch a PIM kernel to memory, as the programmer needs some way to indicate which regions of a program should be offloaded to PIM and which regions should not be offloaded. Our TensorFlow case study in Section 3.3 shows that offloading at the granularity of functions provides speedups of up to 98.1% when the 16 most time- and energy-consuming GEMM operations make use of PIM accelerators for packing and quantization [16]. As we discuss in Sections 4.2 and 4.3, another issue with this approach is the need to coordinate between the CPU and PIM logic, as CPU threads and PIM kernels can potentially execute concurrently. Examples of this coordination include cache coherence [22-24] and address translation [62]. We note that using simple pragmas to indicate the beginning and end of a PIM kernel represents a first step for identifying the blocks of instructions in a program that should be offloaded to PIM, and we encourage future works to develop more robust and expressive interfaces and mechanisms for PIM offloading that can allow for better coordination between the CPU and PIM logic (e.g., by building on expressive memory interfaces [144, 145]).

At the other extreme, a PIM kernel can consist of an *entire application*. Executing an entire application in memory can avoid the need to communicate at all with the CPU. For example, there is no need to perform cache coherence (see Section 4.2) between the CPU and PIM logic, as they work on different programs entirely. While this is a simple solution to maintain programmability and avoid significant modification to hardware, it significantly limits the types of applications that can be executed with PIM. As we discuss in Section 3.2, applications with significant computational complexity or high temporal locality are best suited for the CPU, but significant portions of these applications can benefit from PIM. In order to obtain benefits for such applications when only the entire application can be offloaded, changes must be made across the entire system. We briefly examine two successful examples of entire application offloading: Tesseract [5] and GRIM-Filter [6].

Tesseract [5] is an accelerator for in-memory graph processing. Tesseract adds an in-order core to each vault in an HMC-like 3D-stacked memory, and implements an efficient communication protocol between these in-order cores. Tesseract combines this new architecture with a message-passing-based programming model, where message passing is used to perform operations on the graph nodes by moving the operations to the vaults where the corresponding graph nodes are stored. For five state-of-the-art graph processing workloads with large real-world graphs, Tesseract improves the average system performance by 13.8x, and reduces the energy consumption by 87%, over a conventional CPU-only system [5]. Other recent works build on Tesseract by improving locality and communication for further benefits [146, 147].

GRIM-Filter [6] is an in-memory accelerator for genome seed filtering. In order to read the genome (i.e., DNA sequence) of an organism, geneticists often need to reconstruct the genome from small segments of DNA known as *reads*, as current DNA extraction techniques are unable to extract the entire DNA sequence. A genome *read mapper* can perform the reconstruction by matching the reads against a *reference genome*, and a core part of read mapping is a computationally-expensive dynamic programming algorithm that *aligns* the reads to the reference genome. One technique to significantly improve the performance and efficiency of read mapping is *seed filtering* [148-151], which reduces the number of reference genome *seeds* (i.e., segments) that a read must be checked against for alignment by quickly eliminating seeds with no probability of matching. GRIM-Filter proposes a state-of-the-art filtering algorithm, and places the entire algorithm inside memory [6]. This requires adding simple accelerators in the logic layer of 3D-stacked memory, and introducing a communication protocol between the read mapper and the filter. The communication protocol allows GRIM-Filter to be integrated into a full genome read mapper (e.g., FastHASH [148], mrFAST [152], BWA-MEM [153]), by allowing (1) the read mapper to notify GRIM-Filter about the DRAM addresses on which to execute customized in-memory filtering operations, (2) GRIM-Filter to notify the read mapper once the filter generates a list of seeds for alignment. Across 10 real genome read sets, GRIM-Filter improves the performance of a full state-of-the-art read mapper by 3.65x over a conventional CPU-only system [6].

### 4.2. Sharing Data Between PIM Logic and CPUs

In order to maximize resource utilization within a system capable of PIM, PIM logic should be able to execute at the same time as CPUs, akin to a multithreaded system. In a traditional multithreaded execution model that uses shared memory between threads, writes to memory must be coordinated between multiple cores, to ensure that threads do not operate on stale data values. Due to the per-core caches used in CPUs, this requires that when one core writes data to a memory address, cached copies of the data held within the caches of other cores must be updated or invalidated, which is known as *cache coherence*. Cache coherence involves a protocol that is designed to handle write permissions for each core, invalidations and updates, and arbitration when multiple cores request exclusive access to the same memory address. Within a chip multiprocessor (CMP), the per-core caches can perform coherence actions over a shared interconnect.

Cache coherence is a major system challenge for enabling PIM architectures as general-purpose execution engines. If PIM processing logic is coherent with the processor, the PIM programming model is relatively simple, as it remains similar to conventional shared memory multithreaded programming, which makes PIM architectures easier to adopt in general-purpose systems. Thus, allowing PIM processing logic to maintain such a simple and traditional shared memory programming model can facilitate the widespread adoption of PIM. However, employing traditional fine-grained cache coherence (e.g., a cache-block based MESI protocol [154]) for PIM forces a large number of coherence messages to traverse the narrow memory channel, potentially undoing the benefits of high-bandwidth and low-latency PIM execution. Unfortunately, solutions for coherence proposed by prior PIM works [5, 21, 75] either place some restrictions on the programming model (by eliminating coherence and requiring message-passing-based programming) or limit the performance and energy gains achievable by a PIM architecture.

To preserve traditional programming models and maximize performance and energy gains, we propose a coherence mechanism for PIM called *CoNDA* [22-24], which does *not* need to send a coherence request for every memory access. Instead, as shown in **Figure 7**, CoNDA enables efficient coherence by having the PIM logic (1) *speculatively* acquire coherence permissions for multiple memory operations over a given period of time (which we call *optimistic execution*; ❶ in the figure), (2) *batch* the coherence requests from the multiple memory operations into a set of compressed coherence *signatures* (❷ and ❸), and (3) send the signatures to the CPU to determine whether the speculation violated any coherence semantics. Whenever the CPU receives compressed signatures from the PIM core (e.g., when the PIM kernel finishes), the CPU performs *coherence resolution* (❹), where it checks if any coherence conflicts occurred. If a conflict exists, any dirty cache line in the CPU that caused the conflict is flushed, and the PIM core rolls back and re-executes the code that was optimistically executed. Our execution model shares similarities with Bulk-style mechanisms [155-159] (i.e., mechanisms that speculatively execute chunks of instructions and use speculative information on memory accesses to track potential data conflicts), and with works that use transactional memory (TM) semantics (e.g., [160-164]). However, unlike these past works, the CPU in CoNDA executes *conventionally*, does *not* bundle multiple memory accesses into an atomic transaction, and *never rolls back*, which can make it easier to enable PIM by avoiding the need for complex checkpointing logic or memory access bundling in a sophisticated out-of-order superscalar CPU.

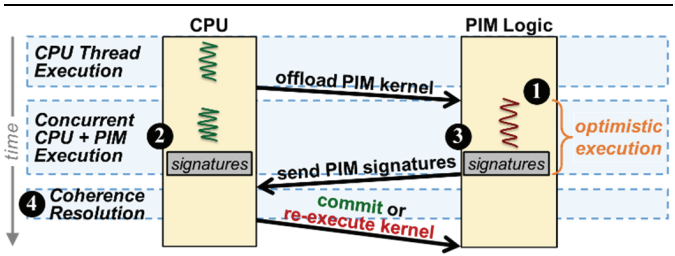

**Figure 7** High-level operation of CoNDA. Adapted from [22].

**Figure 8** shows the performance, normalized to CPU-only, of CoNDA and several state-of-the-art cache coherence mechanisms for PIM [22-24]: FG (fine-grained coherence per cache line), CG (coarse-grained locks on shared data regions), and NC (non-cacheable data regions). We demonstrate that for applications such as graph workloads and HTAP databases, CoNDA improves average performance by 66.0% over the best prior coherence mechanism for performance (FG), and comes within 10.4% of an unrealistic ideal PIM mechanism where coherence takes place instantly with no cost (Ideal-PIM in Figure 8). For the same applications, CoNDA reduces memory system energy by 18.0% (not shown) over the best prior coherence mechanism for memory system energy (CG). CoNDA's benefits increase as application data sets become larger [22]: when we increase the dataset sizes by an order of magnitude (not shown), we find that CoNDA improves performance by 8.4x over CPU-only and by 38.3% over the best prior coherence mechanism for performance (FG), coming within 10.2% of Ideal-PIM.

In our prior work on CoNDA [22-24], we provide a detailed discussion of (1) the need for a new coherence model for workloads such as graph frameworks and HTAP databases, (2) the hardware support needed to enable the CoNDA coherence model, and (3) a comparison of CoNDA to multiple state-of-the-art coherence models.

### 4.3. Virtual Memory

A significant hurdle to efficient PIM execution is the need for virtual memory. An application operates in a virtual address space, and when the application needs to access its data inside main memory, the CPU core must first perform an *address translation*, which converts the data's virtual address into a *physical* address within main memory. The mapping between a virtual address and a physical address is stored in memory in a multi-level *page table*. Looking up a single virtual-to-physical address mapping requires one memory access *per level*, incurring a significant performance penalty. In order to reduce this penalty, a CPU contains a *translation lookaside buffer* (TLB), which caches the most recently used mappings. The CPU also includes a *page table walker*, which traverses the multiple page table levels to retrieve a mapping on a TLB miss.

PIM kernels often need to perform address translation, such as when the code offloaded to memory needs to traverse a pointer. The pointer is stored as a virtual address, and must be translated before PIM logic can access the physical location in memory. A simple solution to provide address translation support for PIM logic could be to issue any translation requests from the PIM logic to the CPU-side virtual memory structures. However, if the PIM logic needs to communicate with existing CPU-side address translation mechanisms, the benefits of PIM could easily be nullified, as each address translation would need to perform a long-latency request across the memory channel. The translation might sometimes require a page table walk, where the CPU must issue multiple memory requests to read the page table, which would further increase traffic on the memory channel.

A naive solution is to simply duplicate the TLB and page walker within memory (i.e., within the PIM logic). Unfortunately, this is prohibitively difficult or expensive for three reasons: (1) coherence would have to be maintained between the CPU and memory-side TLBs, introducing extra complexity and off-chip requests; (2) the duplicated hardware is very costly in terms of storage overhead and complexity; and (3) a memory module can be used in conjunction with many different processor architectures, which use different page table implementations and formats, and ensuring compatibility between the in-memory TLB/page walker and all of these different processor architectures is difficult.

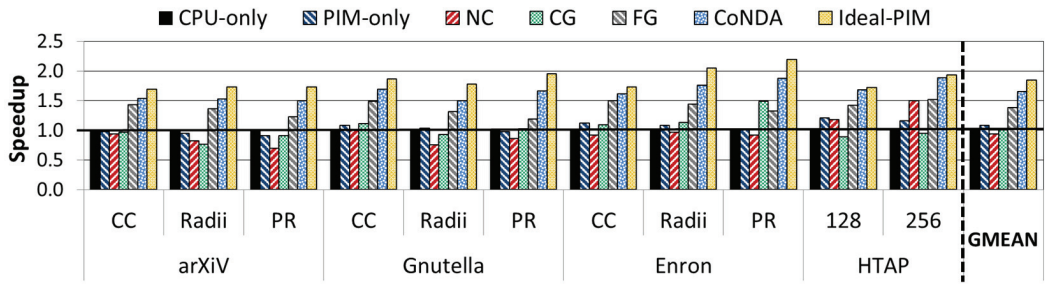

**Figure 8** Speedup of PIM with various cache coherence mechanisms, including CoNDA [22-24], normalized to CPU-only. Adapted from [22].

We study how to solve the challenge of virtual memory translation in the context of IMPICA, our in-memory accelerator for efficient pointer chasing [62]. In order to maintain the performance and efficiency of the PIM logic, we completely decouple the page table of the PIM logic from that of the CPU. This presents us with two advantages. First, the page table logic in PIM is no longer tied to a single architecture (unlike the CPU page table, which is part of the architectural specification), and allows a memory chip with PIM logic to be paired with any CPU. Second, we now have an opportunity to develop a new page table design that is much more efficient for our in-memory accelerator.

We make two key observations about the behavior of a pointer chasing accelerator. First, the accelerator operates only on certain data structures that can be mapped to contiguous regions in the virtual address space, which we refer to as *PIM regions*. As a result, it is possible to map contiguous PIM regions with a *smaller, region-based* page table without needing to duplicate the page table mappings for the entire address space. Second, we observe that if we need to map only PIM regions, we can collapse the hierarchy present in conventional page tables, which allows us to limit the hardware and storage overhead of the PIM page table, and cut the number of memory accesses per page table walk down from four (for a conventional four-level page table) to two. Based on these observations, we build an efficient page table structure that can be used for a wide range of PIM accelerators, where the accelerators access data in only one region of the total data set. The region-based page table improves the performance of IMPICA by 13.5%, averaged across three linked data traversal benchmarks and a real database workload [165], over a conventional four-level page table. Note that these benefits are part of the 34% performance improvement that IMPICA provides across all of the workloads. More detail on our page table design can be found in our prior work on IMPICA [62].

### 4.4. Enabling Programmers and Compilers to Find PIM Targets

In Section 3.2, we discuss our toolflow for identifying PIM targets. While this toolflow is effective at enabling the co-design of PIM architectures and applications that can take advantage of PIM, it still requires a non-trivial amount of effort on the part of the programmer, as the programmer must first run the toolflow, and then annotate programs using directives such as the ones we discuss in Section 4.1. There is a need to develop even easier methodologies for finding PIM targets. One alternative is to automate the toolflow, by developing a PIM compiler that can execute the toolflow and then automatically annotate the portions of an application that should be offloaded to PIM. For example, TOM [75] proposes a compiler-based technique to automatically (1) identify basic blocks in GPU applications that should be offloaded to PIM; and (2) map the data needed by such blocks appropriately to memory modules, so as to minimize data movement. Another alternative is to provide libraries of common functions that incorporate PIM offloading. Programmers could simply call library functions, without worrying about how PIM offloading takes place, allowing them to easily take advantage of the benefits of PIM. There has been little work in this area to date, and we strongly encourage future researchers and developers to explore these approaches to programming PIM architectures.

## 5. Related Work

We briefly survey recent related works in processing-in-memory. We provide a brief discussion of early PIM proposals in Section 2.1.

**Processing-Near-Memory for 3D-Stacked Memories:** With the advent of 3D-stacked memories, we have seen a resurgence of PIM proposals [13, 14, 20, 82]. Recent PIM proposals add compute units within the logic layer to exploit the high bandwidth available. These works primarily focus on the design of the underlying logic that is placed within memory, and in many cases propose special-purpose PIM architectures that cater only to a limited set of applications. These works include accelerators for matrix multiplication [91], data reorganization [117], graph processing [5, 22-24, 84], databases [22-24, 65], in-memory analytics [68], MapReduce [87], genome sequencing [6], data-intensive processing [70], consumer device workloads [16], machine learning workloads [16, 66, 77, 79], and concurrent data structures [81]. Some works propose more generic architectures by adding PIM-enabled instructions [21], GPGPUs [75, 85, 90], single-instruction multiple-data (SIMD) processing units [83], or reconfigurable hardware [67, 69, 71] to the logic layer in 3D-stacked memory. A recently-developed framework [25, 166] allows for the rapid design space exploration of processing-near-memory architectures.

**Processing-Using-Memory:** A number of recent works have examined how to perform memory operations directly within the memory array itself, which we refer to as processing using memory [13, 14, 20, 49]. These works take advantage of inherent architectural properties of memory devices to perform operations in bulk. While such works can significantly improve computational efficiency within memory, they still suffer from many of the same programmability and adoption challenges that PIM architectures face, such as the address translation and cache coherence challenges that we focus on in this article. Mechanisms for processing-using-memory can perform a variety of functions, such as bulk copy and data initialization for DRAM [17, 19]; bulk bitwise operations for DRAM [18, 47-49, 80, 118], PCM [41], or MRAM [44-46]; and simple arithmetic operations for SRAM [63, 76, 100] and RRAM/memristors [42, 43, 88, 92-99].

**Processing in the DRAM Module or Memory Controller:** Several works have examined how to embed processing functionality near memory, but not within the DRAM chip itself. Such an approach can reduce the cost of PIM manufacturing, as the DRAM chip does not need to be modified or specialized for any particular functionality. However, these works (1) are often unable to take advantage of the high internal bandwidth of 3D-stacked DRAM, which reduces the efficiency of PIM execution, and (2) may still suffer from many of the same challenges faced by architectures that embed logic within the DRAM chip. Examples of this work include (1) Gather-Scatter DRAM [87], which embeds logic within the memory controller to remap a single memory request across multiple rows and columns within DRAM; (2) work by Hashemi et al. [72, 73] to embed logic in the memory controller that accelerates dependent cache misses and performs runahead execution [167]; and (3) Chameleon [64] and the Memory Channel Network architecture [168], which propose methods to integrate logic within the DRAM module but outside of the chip to reduce manufacturing costs.

**Addressing Challenges to PIM Adoption:** Recent work has examined design challenges for systems with PIM support that can affect PIM adoption. A number of these works improve PIM programmability, such as CoNDA [22-24], which provides efficient cache coherence support for PIM; the study by Sura et al. [89], which optimizes how programs access PIM data; PEI [21], which introduces an instruction-level interface for PIM that preserves the existing sequential programming models and abstractions for virtual memory and coherence; TOM [75], which automates the identification of basic blocks that should be offloaded to PIM and the data mapping for such blocks; work by Pattnaik et al. [85], which automates whether portions of GPU applications should be scheduled to run on GPU cores or PIM cores; and work by Liu et al. [81], which designs PIM-specific concurrent data structures to improve PIM performance. Other works tackle hardware-level design challenges, including IMPICA [62], which introduces in-memory support for address translation and pointer chasing; work by Hassan et al. [74] to optimize the 3D-stacked DRAM architecture for PIM; and work by Kim et al. [78] that enables PIM logic to efficiently access data across multiple memory stacks. There is recent work on modeling and understanding the interaction between programs and PIM hardware, such as NAPEL [25, 166], a framework that predicts the potential performance and energy benefits of using PIM.

## 6. Future Challenges

In Sections 3 and 4, we demonstrate the need for several solutions to ease programming effort in order to take advantage of the benefits of PIM. We believe that a number of other challenges remain for the widespread adoption of PIM:

- *PIM Programming Model:* Programmers need a well-defined interface to incorporate PIM functionality into their applications. While we briefly discuss several interfaces and mechanisms for offloading different granularities of applications to PIM, defining a complete programming model for how a programmer should invoke and interact with PIM logic remains an open problem.
- *Data and Logic Mapping:* To maximize the benefits of PIM, all of the data that needs to be read from or written to by a single PIM kernel or by a single PIM core should be mapped to the same memory stack or memory channel [21, 75]. This requires system architects and programmers to rethink how and where data is allocated. Likewise, for processing-using-memory architectures, programs often require more complex logic functions than the bitwise operations enabled by these architectures, and require some form of logic mapping or synthesis to allow programmers or compilers to efficiently implement these more complex logic functions on the processing-using-memory substrates [169-171]. There is a need to develop robust programmer-transparent data mapping and logic mapping/synthesis support for PIM architectures.
- *PIM Runtime Scheduling:* There needs to be coordination between PIM logic and the PIM kernels that are either being executed currently or waiting to be executed. Determining when to enable and disable PIM execution [21], what to execute in memory, how to share PIM cores and PIM accelerators across multiple CPU threads/cores, and how to coordinate between PIM logic accesses and CPU accesses to memory are all important runtime attributes that must be addressed.

New performance and energy prediction frameworks [25] and simulation tools [166] can help researchers with solving several of the remaining challenges. We refer the reader to our overview works [13, 14, 172, 173] on enabling the adoption of PIM for further discussion of these challenges.

## 7. Conclusion

While many important classes of emerging AI, machine learning, and data analytics applications are operating on very large data sets, conventional computer systems are not designed to handle such large-scale data. As a result, the performance and energy costs associated with moving data between main memory and the CPU dominate the total costs of computation, which is a phenomenon known as the *data movement bottleneck*. To alleviate this bottleneck, a number of recent works propose *processing-in-memory* (PIM), where unnecessary data movement is reduced or eliminated by bringing some or all of the computation into memory. There are many practical system-level challenges that need to be solved to enable the widespread adoption of PIM. In this work, we examine how these challenges relate to programmers and system architects, and describe several of our solutions to facilitate the systematic offloading of computation to PIM logic. In a case study, we demonstrate our offloading toolflow with Google's TensorFlow Lite framework for neural network inference, demonstrating that we can achieve performance improvements of up to 98.1%, while reducing energy consumption by an average of 54.9%. We then discuss the need for mechanisms that preserve conventional programming models when offloading computation to PIM. We discuss several such mechanisms, which provide various methods of offloading portions of applications to PIM logic, sharing data between PIM logic and CPUs, enabling efficient virtual memory access for PIM, and automating PIM target identification and offloading. Finally, we describe a number of remaining challenges to the widespread adoption of PIM. We hope that our work and analysis inspire researchers to tackle these remaining challenges, which can enable the commercialization of PIM architectures.

**Acknowledgments**
We thank all of the members of the SAFARI Research Group, and our collaborators at Carnegie Mellon, ETH Zürich, and other universities, who have contributed to the various works we describe in this article. Thanks also goes to our research group's industrial sponsors over the past ten years, especially Alibaba, Facebook, Google, Huawei, Intel, Microsoft, NVIDIA, Samsung, Seagate, and VMware. This work was also partially supported by the Intel Science and Technology Center for Cloud Computing, the Semiconductor Research Corporation, the Data Storage Systems Center at Carnegie Mellon University, past NSF grants 1212962, 1320531, and 1409723, and past NIH grant HG006004.

**Saugata Ghose** *Carnegie Mellon University, Pittsburgh, PA, USA (ghose@cmu.edu)*. Dr. Ghose received dual B.S. degrees in Computer Science and in Computer Engineering from Binghamton University, State University of New York, in 2007, and received an M.S. degree and a Ph.D. degree in Computer Engineering from Cornell University in 2014. Since 2014, he has been working at Carnegie Mellon University, where he is currently a systems scientist in the department of Electrical and Computer Engineering. Dr. Ghose was the recipient of the NDSEG Fellowship and the ECE Director's Ph.D. Teaching Award while at Cornell, received the best paper award at the 2017 DFRWS Digital Forensics Research Conference Europe, and won a Wimmer Faculty Fellowship at CMU. His current research interests include processing-in-memory, low-power memories, application-and system-aware memory and storage systems, and data-driven architectures. For more information, see his webpage at https://ece.cmu.edu/~saugatag/

**Amirali Boroumand** *Carnegie Mellon University, Pittsburgh, PA, USA (amirali@cmu.edu)*. Mr. Boroumand received a B.S. degree in Computer Hardware Engineering from the Sharif University of Technology in 2014. Since 2014, he has been a Ph.D. student at Carnegie Mellon University. His current research interests include programming support for processing-in-memory, and in-memory architectures for consumer devices and for databases.

**Jeremie S. Kim** *Carnegie Mellon University, Pittsburgh, PA, USA; ETH Zürich, Zürich, Switzerland (jeremiek@andrew.cmu.edu)*. Mr. Kim received a B.S. degree and an M.S. degree in Electrical and Computer Engineering from Carnegie Mellon University in 2015. He is currently working on his Ph.D. with Onur Mutlu at Carnegie Mellon University and ETH Zürich. His current research interests are in computer architecture, memory latency/power/reliability, hardware security, and bioinformatics, and he has several publications on these topics.

**Juan Gómez-Luna** *ETH Zürich, Zürich, Switzerland (juang@ethz.ch)*. Dr. Gómez-Luna received a B.S. and an M.S. degree in Telecommunication Engineering from the University of Seville in 2001, and received a Ph.D. degree in Computer Science from the University of Córdoba in 2012. Between 2005 and 2017, he was a lecturer at the University of Córdoba. Since 2017, Dr. Gómez-Luna has been a postdoctoral researcher at ETH Zürich. His current research interests focus on software optimization for GPUs and heterogeneous systems, GPU architectures, near memory processing, medical imaging, and bioinformatics.

**Onur Mutlu** *ETH Zürich, Zürich, Switzerland; Carnegie Mellon University, Pittsburgh, PA, USA (omutlu@ethz.ch)*. Dr. Mutlu received dual B.S. degrees in Computer Engineering and in Psychology from the University of Michigan in 2000, and an M.S. degree in 2002 and a Ph.D. degree in 2006 in Electrical and Computer Engineering from the University of Texas at Austin. He is currently a professor of Computer Science at ETH Zürich. He is also a faculty member at Carnegie Mellon University, where he previously held the William D. and Nancy W. Strecker Early Career Professorship. Dr. Mutlu's industrial experience includes starting the Computer Architecture Group at Microsoft Research, where he worked from 2006 to 2009, and various product and research positions at Intel Corporation, Advanced Micro Devices, VMware, and Google. He received the ACM SIGARCH Maurice Wilkes Award, the inaugural IEEE Computer Society Young Computer Architect Award, the inaugural Intel Early Career Faculty Award, faculty partnership awards from various companies, and a healthy number of best paper and "Top Pick" paper recognitions at various computer systems and architecture venues. He is an ACM Fellow, IEEE Fellow, and an elected member of the Academy of Europe (Academia Europaea). His current broader research interests are in computer architecture, systems, and bioinformatics. He is especially interested in interactions across domains and between applications, system software, compilers, and microarchitecture, with a major current focus on memory and storage systems. For more information, see his webpage at https://people.inf.ethz.ch/omutlu/